\documentclass[letterpaper,twocolumn]{jpsj3}
\usepackage{txfonts}

\title{An Approximation Method for the Scattering Data of One--Dimensional Soliton Equations under Arbitrary Rapidly Decreasing Initial Pulses}

\author{Hironobu Fujishima$^1$\thanks{fujishima@lamp.is.utsunomiya-u.ac.jp} and Tetsu Yajima$^2$\thanks{yajimat@is.utsunomiya-u.ac.jp}}
\inst{$^1$Utsunomiya Optical Products Plant, Canon Inc., 20-2 Kiyohara Kogyoudanchi,
Utsunomiya, Tochigi 321-3231, Japan \\
$^2$Department of Information Systems Science, Graduate School of
Engineering, Utsunomiya University, 7-1-2 Yoto, Utsunomiya, 
Tochigi 321-8585, Japan} 

\abst{We present a novel approximation method that can predict the number of solitons asymptotically appearing under arbitrary rapidly decreasing initial wave packets. The number of solitons can be estimated without integration of the original soliton equations. As an example, we take the one--dimensional nonlinear Schr\"odinger equation and estimate the behaviors of the scattering amplitude in detail. The results show good agreement compared with those obtained by direct numerical integration. The presented method is applicable to a wide class of one--dimensional soliton equations.
}

\begin{document}
\maketitle

\section{Introduction}
Soliton theory is one of the key concepts in mathematical physics. It is particularly important for continuous systems owing to the fact that a number of intriguing phenomena across various fields of physics can be systematically described in this common language of solitons. The fields where 
solitons play prominent roles include high energy physics\cite{ref:Rajaraman,ref:Manton,ref:Shifman,ref:Dunajski,ref:Weinberg} and gravitational physics\cite{ref:Belinski}. As examples of  physics in more easily accessible energy scales, we can cite continuum mechanics of fluids or plasmas\cite{ref:Whitham,ref:Karpman,ref:Lamb,ref:Infeld,ref:Dauxois,ref:Ablowitz2} and nonlinear optics\cite{ref:Hasegawa,ref:Abdullaev,ref:Agrawal,ref:Kivshar,ref:Taylor}. In the context of condensed matter or low temperature physics, the macroscopic dynamics of Bose--Einstein condensed (BEC) systems is known to obey a kind of nonlinear Sch\"rodinger equation (NLSE) with external potentials\cite{ref:Griffin,ref:Pitaevskii,ref:Pethick,ref:Ueda}. These rich applications means that this mature subject is still an active area of research. 

Soliton equations are nonlinear partial differential equations often discussed in the context of integrable systems. Although some subtleties exist in rigorously defining integrability for systems with infinite degrees of freedom, it is sufficient for our purpose to point out the applicability of the inverse scattering transformation (IST) method. This prominent property is shared by all soliton equations\cite{ref:Ablowitz1,ref:Calogero,ref:Novikov,ref:Newell,ref:Faddeev,ref:Clarkson}.

 By virtue of the IST method, soliton equations are decomposed into a pair of linear simultaneous differential equations in the following manner:
\begin{subequations}
\label{eq:laxeq}
\begin{align}
\label{eq:laxeq_space}
\Psi_x&=S\Psi,\\
\label{eq:laxeq_time}
\Psi_t&=T\Psi,
\end{align}
\end{subequations}
where the quantities $S$ and $T$ are matrices or operators including the unknown functions of the original equation and an additional parameter called the spectral parameter. The wave function $\Psi$ represents an auxiliary field satisfying appropriate boundary conditions. Equation (\ref{eq:laxeq_space}) takes the form of an eigenvalue problem or a stable scattering problem where the initial condition plays the role of the scattering potential. Equation (\ref{eq:laxeq_time}) governs the time evolution of the scattering data whose initial values are determined by Eq.~(\ref{eq:laxeq_space}).

Utilizing the obtained scattering data, we can construct exact solutions to the original equations at an arbitrary time. It is known that the original initial value problems can be solved under a fairly wide class of initial conditions. A class of rapidly decreasing functions is one of the most important classes of initial conditions for many soliton equations. Needless to say, it is almost impossible to prepare ideal pure solitons in actual experiments and hence it is more realistic to consider more general initial conditions rather than pure solitons. For many soliton equations, initial value functions should be rapidly decreasing, and this should be the only requirement that we can reasonably impose. One of the important quantities characterizing the initial conditions is the number of remaining solitons at $t\to\infty$. Since our target time domain lies in the asymptotic future and because of the diffusing nature of wave packets, the need for subtle treatment of boundary conditions arises. Therefore, a theoretical framework not depending on direct numerical integration will be useful and can serve to endorse the validity of numerical calculations. 

To this end, we introduce a novel approximation method that can predict the number of solitons in the final state without direct numerical integration. The purpose of this paper is to present this method with some illustrative applications. This method provides a way to approximately obtain the scattering data for Eq.~(\ref{eq:laxeq_space}) by discretizing the space coordinate and it is different from the method we used in our previous work\cite{ref:Fujishima}. The method  presented in this paper is intended to overcome the drawbacks of our previously adopted semi--analytic method.

In contrast to the above treatments, the new approximation method presented in this paper is purely numerical and can be applied to realistic smooth functions for which sufficient fineness of the discretization pitch is required.

This paper is organized as follows. In sect.2, we present mathematical preliminaries by taking the NLSE as an example. Section 3 is the central part of this paper and we develop the method for detailed analysis of the initial value problem. In sec.4, we show some illustrative applications and present a brief model of a collision experiment with attractively interacting BEC solitons\cite{ref:Nguyen,ref:Parker,ref:Billam}. We compare the results obtained by the presented method with those from direct numerical integration to check the consistency between them. The final section is devoted to a discussions and concluding remarks. 

\section{Mathematical Preliminaries}
\subsection{Inverse scattering transformation}
To be specific, we consider the NLSE with a self--focusing interaction throughout this paper,
\begin{equation}
i\psi_t=-\psi_{xx}-2|\psi|^2\psi.\label{NLSE}
\end{equation}
We assume that the initial wave packet is a rapidly decreasing function of $x$ hereafter.
The auxiliary linear equations (\ref{eq:laxeq}) for the NLSE are derived by introducing the two--element vector function $\Psi$ and 2$\times$2 matrices $S$ and $T$ as
\begin{subequations}
\label{eq:nlslax_matrices}
\begin{align}
\label{eq:nlslax_space}
S&=
\begin{pmatrix}
-i\xi&i\psi^*\\
i\psi&i\xi
\end{pmatrix},\\
\label{eq:nlslax_time}
T&=
\begin{pmatrix}
2i\xi^2-i|\psi|^2&\psi^*_x-2i\xi\psi^*\\
-\psi_x-2i\xi\psi&-2i\xi^2+i|\psi|^2
\end{pmatrix},
\end{align}
\end{subequations}
where $\xi$ is the time--independent spectral parameter. Many soliton equations can be formulated using the 2$\times$2 matrices in the framework of the IST method and they are classified by the way the spectral parameter $\xi$ appears in these matrices. The NLSE belongs to the Ablowitz--Kaup--Newell--Segur (AKNS) system\cite{ref:AKNS}, and other soliton equations belonging to the AKNS system can be discussed in a similar way. 
\par
An important step of the IST method is to analyze the spatial Eq.~(\ref{eq:laxeq_space}) as a stable scattering problem, whose potential term is given by the initial condition $\psi(x,0)$. The wave function $\Psi$ and the spectral parameter $\xi$ correspond to the eigenfunction and eigenvalue, respectively. 
We assume the rapidly decreasing boundary condition for $\psi(x,0)$ as already mentioned:
\begin{equation}
 \psi(x,0)\to0,\quad\mbox{as $|x|\to\infty$.}
\end{equation}
Equations (\ref{eq:laxeq_space}) and (\ref{eq:nlslax_space}) under the above boundary condition completely define the Zakharov--Shabat (ZS) problem\cite{ref:Zakharov} for the self--focusing NLS equation. 
With this boundary condition, each element of the wave function $\Psi$ must become a plane wave in the limit of $x\to\pm\infty$. As the fundamental 
solutions, we can select two sets of functions ${\{\phi,\bar{\phi}\}}$ and ${\{\chi,\bar{\chi}\}}$ called the Jost 
functions, which satisfy the boundary conditions
\begin{subequations} 
\label{eq:jost_boundaries}
\begin{align}
\label{eq:jost_neginftyx}
 &\phi(x;\xi)\to\begin{pmatrix}e^{-i\xi x}\\0\end{pmatrix},\ 
 \bar\phi(x;\xi)\to\begin{pmatrix}0\\e^{i\xi x}\end{pmatrix},
 &\mbox{as $x\to-\infty$},\\
\label{eq:jost_posinftyx}
 &\chi(x;\xi)\to\begin{pmatrix}0\\e^{i\xi x}\end{pmatrix},\ 
 \bar\chi(x;\xi)\to\begin{pmatrix}e^{-i\xi x}\\0\end{pmatrix},
 &\mbox{as $x\to+\infty$}.
\end{align}
\end{subequations}
The Jost functions are related to each other as
\begin{equation}\label{eq:scattering_rels}
\begin{split}
\phi(x;\xi)=a(\xi)\bar\chi(x;\xi)+b(\xi)\chi(x;\xi),\\
\bar\phi(x;\xi)=\bar a(\xi)\chi(x;\xi)-\bar b(\xi)\bar\chi(x;\xi).
\end{split}
\end{equation}
The coefficients are members of the scattering data, and $a(\xi)$ can be analytically continued to the upper 
half plane $\mathop{\rm Im}\xi>0$.
\par
From Eqs.~(\ref{eq:jost_boundaries}) and (\ref{eq:scattering_rels}),
we can see that the Jost function $\phi(x;\xi)$ satisfies the asymptotic form
\begin{equation}
\label{sd}
\phi(x;\xi)=\begin{pmatrix}a(\xi)e^{-i\xi x}\\
b(\xi)e^{i\xi x}\\
\end{pmatrix}\quad\mbox{as $x\to+\infty$.}.
\end{equation}
Once the ZS problem is solved, the time--evolved wave function is easily obtained by Eq.~(\ref{eq:laxeq_time}), and the solution under the initial value is formally constructed by virtue of the Gel'fand--Levitan--Malchenko (GLM) equation\cite{ref:Gel'fand}:
\begin{equation}
K_{1}^{\ast}(x,y;t)-F(x+y;t)-\int_{x}^{\infty}K_{2}(x,z;t)F(z+y;t)dz=0,
\end{equation}
\begin{equation}
K_{2}^{\ast}(x,y;t)+\int_{x}^{\infty}K_{1}(x,z;t)F(z+y;t)dz=0,
\end{equation}
where the scattering data appear through the function
\begin{equation}
F(x;t)=\frac{1}{2\pi}\int_{-\infty}^{\infty}\frac{b(\xi)}{a(\xi)}e^{i\xi x-4i\xi^{2}t}d\xi-i\sum_{l=1}^{M}\frac{b(\zeta_{l})}{a'(\zeta_{l})}e^{i\zeta_{l}x-4i\zeta_{l}t}.
\end{equation}
The solution of the NLSE is given by the kernel function $K_{1}$ as
\begin{equation}
\psi(x,t)=-2iK_{1}^{\ast}(x,x;t).
\end{equation}

In the above equation, $\xi$ stands for the continuous real eigenvalue and $\eta$ are the discrete eigenvalues, which are the zeros of $a(\xi)$. 
The GLM equation clearly states that the number 
of discrete eigenvalues determines the number of asymptotically generated solitons. When the function $a(\xi)$ has $M$ simple zeros $\xi=\xi_1,\xi_2,\ldots,\xi_M$
on the upper half plane of $\xi$, $M$ solitons appear asymptotically and each $\xi$ determines the characteristics of each soliton.

\subsection{Discretization of the initial wave packet}
As explained in sect.2.1, we need to know $a(\xi)$ to extract the desired information of the final state. We find that this is equivalent to calculating $\phi(x;\xi)$ at $x\to+\infty$ under the boundary condition Eq.~(\ref{eq:jost_neginftyx}).

In the ZS problem for the NLSE
\begin{equation}\label{eq:ZS}
\Psi_x=S\Psi,\quad
S=\begin{pmatrix}
-i\xi&i\psi^\ast\\i\psi&i\xi
\end{pmatrix},
\end{equation}
the function $\psi$ is the initial value of the unknown function $\psi(x,t)$. The ZS problem for the NLSE under the initial condition $\psi(x,0)=A\mathrm{sech}(x)$, where $A$ is a constant, has been extensively studied\cite{ref:Satsuma}. The explicit forms of the wave function $\Psi$ are written by the hypergeometric function and the number of remaining solitons is given by the maximum integer $n$ that satisfies $n\leq A$. Nevertheless, the ZS problems for more general initial conditions are difficult to solve, and it is hardly possible to predict how many solitons remain in the final state. 
\par
The major difficulty in analyzing Eq.~(\ref{eq:ZS}) for general initial conditions originates from the fact that $\psi
(x,0)$ depends on the coordinate $x$. In order to deal with this difficulty, let us split the support of $\psi(x,0)$ into many small intervals:
\begin{equation}
\label{eq:disc}
I_j:\ x_j\le x<x_{j+1}\quad (j=1,\ldots,N),
\end{equation}
and approximate $\psi(x,0)$ so that it takes a constant value in each interval. We introduce a set of functions $
\psi_j$:
\begin{equation}
\psi_j(x)=
\begin{cases}
V_j&x\in I_j,\\
0&x\notin I_j.  \end{cases}
\end{equation}
The initial value $\psi(x,0)$ is now approximated as
\begin{align}
\label{eq:discV}
\psi(x,0)&=\sum_{j=1}^N\psi_j(x),\\
&=
\begin{cases}
V_j&(x\in I_j,\ j=1,2,\ldots,N),\\
0&\mbox{(otherwise).}
\end{cases}
\end{align}
Assuming $\psi(x,0)$ belongs to the class of rapidly decreasing functions, 
we can approximately consider that $\psi(x,0)$ has a compact support.
Within each interval, Eq.~(\ref{eq:ZS}) reads as
\begin{equation}\label{eq:splitted_lax}
\Psi_x=S_j\Psi,\quad
S_j=\begin{pmatrix}
-i\xi&iV^\ast_j\\iV_j&i\xi
\end{pmatrix}.
\end{equation}

\section{New Approximation Method}
Now, let us develop our method with the discretization scheme. First, we introduce a two--component vector function $\Psi$:
\begin{equation}
\Psi(x,\xi)=
\begin{pmatrix}
\Psi_{1}(x,\xi) \\
\Psi_{2}(x,\xi)
\end{pmatrix}, \quad
\Psi_{0}(x,\xi)=
\begin{pmatrix}
e^{-i\xi x} \\
0
\end{pmatrix}.
\end{equation}
The 2$\times$2 matrix Green function $g(x-x')$ is defined using the step function $\theta(x)$ as
\begin{equation}
g(x-x')=
\begin{pmatrix}
e^{-i\xi(x-x')}\theta(x-x')&0\\
0&e^{i\xi(x-x')}\theta(x-x')
\end{pmatrix},
\label{eq:Green}
\end{equation}
and the scattering potential in the matrix form is denoted by
\begin{equation}
U(x)=
\begin{pmatrix}
0&i\psi^{\ast}(x,0)\\
i\psi(x,0)&0
\end{pmatrix}.
\end{equation}
The starting point is to reformulate the ZS eigenvalue problem in Eq.~(\ref{eq:ZS}) as Fredholm integral equations of the second kind: 
\begin{equation}
\label{eq:integral}
\Psi(x,\xi)=
\Psi_{0}(x,\xi)
+\int^{\infty}_{-\infty} dx'
g(x-x')U(x')\Psi(x',\xi).
\end{equation}
Needless to say, the solution $\Psi(x,\xi)$ asymptotically coincides with the Jost function in Eq.~(\ref{sd}) at the limit $x\to+\infty$.
By iteration, we can construct the perturbative solution to Eq.~(\ref{eq:integral}). By considering the first element of the two--component vector $\Psi(x,\xi)$, we can write down following Neumann series:
\begin{equation}
\begin{split}
\Psi_{1}(x,\xi)=e^{-i\xi x}+&e^{-i\xi x}\int_{-\infty}^{\infty}dx'dx''\theta(x-x')e^{2i\xi x'}i\psi^{\ast}(x',0)\\&\cdot\theta(x'-x'')e^{-2i\xi x''}i\psi(x'',0)+\mathrm{(even)}.
\end{split}
\end{equation}
According to Eq.~(\ref{sd}), the transmission coefficient $a(\xi)$ can be obtained by taking large--$x$ limit:
\begin{equation}
a(\xi)e^{-i\xi x}=\lim_{x\to\infty}\Psi_{1}(x,\xi)=\lim_{x\to\infty}\phi_{1}(x,\xi).
\end{equation}
Note that only terms of even orders contribute to the transmission coefficient $a(\xi)$. This is because the number of iterative operations is equal to the number of scattering events that invert the direction of the incident plane wave $\Psi_{0}(x,\xi)$.

Similarly, by considering the second element of the two--component vector $\Psi(x,\xi)$, we can obtain the reflection coefficient $b(\xi)$, to which only terms of odd order contribute. 

For later use, it is convenient to redefine the potential term to include the phase factor $e^{2i\xi x}$,  
 \begin{equation}
v(x)=ie^{2i\xi x}\psi^{\ast}(x,0), 
\label{eq:potential}
\end{equation}
\begin{equation}
v^{\ast}(x)=ie^{-2i\xi x}\psi(x,0).
\end{equation}
By using the above notations, the expression for the transmission coefficient $a(\xi)$ can be rewritten as
\begin{subequations}
\label{eq:aint}
\begin{align}
\label{eq:alim}
a(\xi)=1+\lim_{x\to\infty}A(x;\xi),
\end{align}
\begin{align}
\begin{split}
A(x;\xi)=\int_{-\infty}^{\infty}dx'\int_{-\infty}^{\infty}dx''&\theta(x-x')v(x')\\&\cdot\theta(x'-x'')v^{\ast}(x'')+\mathrm{(even)}.
\end{split}
\end{align}
\end{subequations}

We now consider the replacement of the integral in Eq.~ (\ref{eq:aint}) by a matrix operation of finite size. Although the reduced expression for $a(\xi)$ becomes an approximate one, it is more suitable for numerical calculation. We assume that the initial wave packet is distributed in a finite range of width $L$. In order to carry out the integral and the limit in Eq.~(\ref{eq:aint}), we use the discrete variables $x_j$ ($j=1,2,\ldots,N+1$) defined in Eq.~(\ref{eq:disc}), and we have the following approximate expression for $a(\xi)$:
\begin{equation}
\label{eq:alimn}
a(\xi)=1+\lim_{n\to\infty}A(x_n;\xi).
\end{equation}
By using these discrete variables, the value of $A(x;\xi)$ at discrete values of $x$ can be expressed by a finite sum as
\begin{equation}
A(x_j;\xi)=\sum_{k,l=1}^{N}\theta(x_j-x_k)v(x_k)\theta(x_k-x_l)v^*(x_l)(\Delta x)^2+(\textrm{even}),
\end{equation}
where $\Delta x$ is the spacing of the discrete variables, given as $\Delta x=L/N$.
Since the quantites $v(x_k)$ and $v^*(x_l)$ are defined in the intervals in Eq.~(\ref{eq:disc}), they are expressed by $V_j$ in Eq.~(\ref{eq:discV}) as
\begin{equation}
v(x_k)=ie^{2i\xi x_k}V^*_k,\qquad
v^*(x_l)=ie^{-2i\xi x_k}V_l.
\end{equation}
We introduce matrices $V$ and $V^*$ as
\begin{align}
\begin{aligned}
V&=\mathop{\mathrm{diag}}(ie^{2i\xi x_1}V_1^*,\ldots,ie^{2i\xi x_n}V_N^*),\\
V^*&=\mathop{\mathrm{diag}}(ie^{-2i\xi x_1}V_1,\ldots,ie^{2i\xi x_n}V_N).\end{aligned}
\end{align}
We also define a step function matrix $G$ whose $(i,j)$--element is given by $\theta(x_i-x_j)\Delta x$. The matrix $G$ is a lower--triangular matrix, and its explicit form is
\begin{equation}
G=\dfrac{L}N\begin{pmatrix}
1&0&0&\cdots&0\\
1&1&0&\cdots&0\\
1&1&1&&0\\
\vdots&\vdots&\vdots&\ddots&\vdots\\
1&1&1&\cdots&1
\end{pmatrix}.
\end{equation}
The size of the initial wave packet should be sufficiently large so that the rapidly decreasing tails of the initial wave packet become negligible. To meet the convergence condition for the Neumann series, we can choose a sufficiently large number of small intervals so that $|V_{i}|L/N$ is always smaller than unity for every index $i$. 

From the matrices $V$, $V^*$, and $G$, we find that
\begin{equation}
A(x_j;\xi)=\sum_{l=1}^N(GVGV^*)_{jl}+(\textrm{even}).
\end{equation}
This means that $A(x_j;\xi)$ is given by the sum of the elements in the $j$th row of the matrix
\begin{equation}
W=GVGV^*+(GVGV^*)^2+\cdots.
\end{equation}
If we introduce the column vector $\mbox{\boldmath{$e$}}=(1,\ldots,1)^{\rm T}$ and write the $j$th row of $W$ as $\mbox{\boldmath{$w$}}_j$, we obtain
\begin{equation}
A(x_j;\xi)=\mbox{\boldmath{$w$}}_j\cdot\mbox{\boldmath{$e$}}.
\end{equation}
Since the scattering coefficient is given by Eq.~(\ref{eq:alimn}) and the initial pulse is located in the region $x\le x_{N+1}$, we can derive an approximate expression for $a(\xi)$ as
\begin{equation}
a(\xi)=1+A(x_N;\xi).
\end{equation}
This is equivalent to the expression
\begin{equation}
\label{eq:result}
a(\xi)=\sum_{k=1}^{N}\left(\frac{1}{1-GVGV^{\ast}}\right)_{N,k}
\begin{pmatrix}
1 \\
\vdots \\
1
\end{pmatrix}_{k}\\.
\end{equation}
Many useful algorithms for obtaining the inverse matrix are available and we can calculate Eq.~(\ref{eq:result}) easily.

Theoretically, since the fact the matrix product of $G$ and $V$ is still a lower--triangular matrix, we can explicitly write down the formula for $a(\xi)$ as
\begin{equation}
a(\xi)=\frac{1}{C_{11}}-\left(\frac{C_{12}}{C_{11}C_{22}}+\frac{C_{13}}{C_{11}C_{13}}+\frac{C_{23}}{C_{22}C_{33}}+\cdots\right)+\cdots,
\end{equation}
where the following notations are used:
\begin{equation}
\label{eq:A}
C_{i,j}=[1-GVGV^{\ast}]_{i,j}=\delta_{ij}-\frac{L^{2}}{N^{2}}\sum_{l=j}^{i}V_{l}V^{\ast}_{j}.
\end{equation}
Although the factor $1/N^{2}$ exists in Eq.~(\ref{eq:A}), the summation takes values between the first order and second order of the quantity $1/N$ as a whole.

\section{Applications}
We show some specific applications of the presented method. First, we calculate the transmission coefficient $a(\xi)$ for two different types of single--peak initial conditions, both of which are rapidly decreasing smooth functions. Through this simple example, the validity and effectiveness of our method is corroborated. Secondly, we consider two superposed wave packets whose centers are sufficiently separated so that the resultant initial packet forms a double--peaked profile. The phases of the two constituent pulses are varied to investigate the effect of the relative phases on the asymptotically generated states. The prediction obtained from our time--independent method is compared with the result from direct numerical integration of the original NLSE. 

\subsection{Sech--type and Gauss--type wave packets}
\begin{figure}
\includegraphics{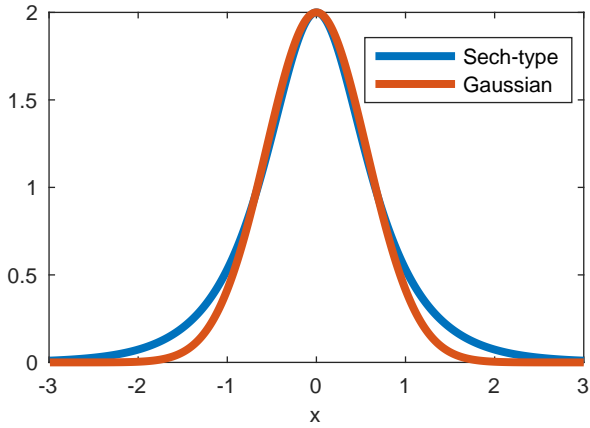}
\caption{(Color online) Two kinds of typical initial wave packets given in Eqs.~(\ref{eq:sech}) and (\ref{eq:Gauss}).}
\label{F1}
\end{figure}

We consider two typical initial conditions for the NLSE in Eq.~(\ref{NLSE}), one of which is taken to be the pure one--soliton
\begin{subequations}
\label{eq:initWFs}
\begin{equation}
\label{eq:sech}
\psi_{\mathrm{s}}(x)=2\mathrm{sech}(x),
\end{equation}
and the other is the Gaussian wave packet
\begin{equation}
\label{eq:Gauss}
\psi_{\mathrm{G}}(x)=2e^{-\frac{\pi}{2}x^{2}}.
\end{equation}
\end{subequations}
Both functions are shown in Fig.~\ref{F1}. Their profiles are similar. They also share the same $L^{2}$--norm and have almost the same full width at half maximum (FWHM). Within the FWHM, the difference between these two functions is no more than 4\% of their maximum value of 2 at the origin.
We take the values of $L$ and $N$ of $-3.0\leq L\leq 3.0$ and $N=1200$ and all the tails of these functions are truncated at $x=\pm3.0$. We calculate the transmission coefficient $a(\xi)$ for these truncated wave packets using the method developed in the previous section.

\begin{figure}
\includegraphics{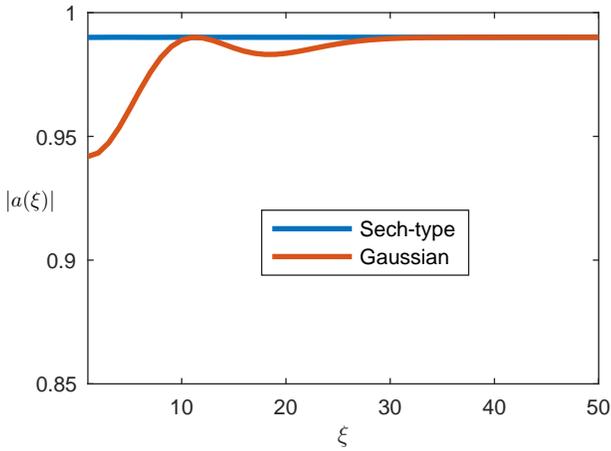}
\caption{(Color online) Absolute values of $a(\xi)$ for typical initial wave packets given by Eqs.~(\ref{eq:sech}) and (\ref{eq:Gauss}). The imaginary part of $\xi$ is set to be vanishing.}
\label{F2}
\end{figure}

\begin{figure}
\includegraphics{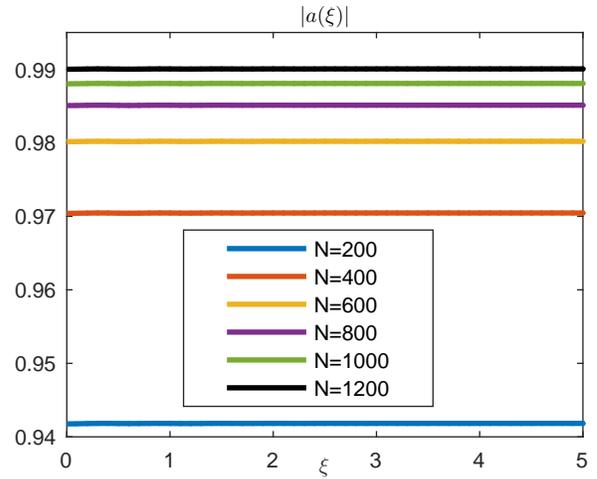}
\caption{(Color online) Absolute values of $a(\xi)$ for various $N$. These plots were calculated for the same initial wave packet Eq.~(\ref{eq:sech}).}
\label{F3}
\end{figure}

\begin{figure}
\includegraphics{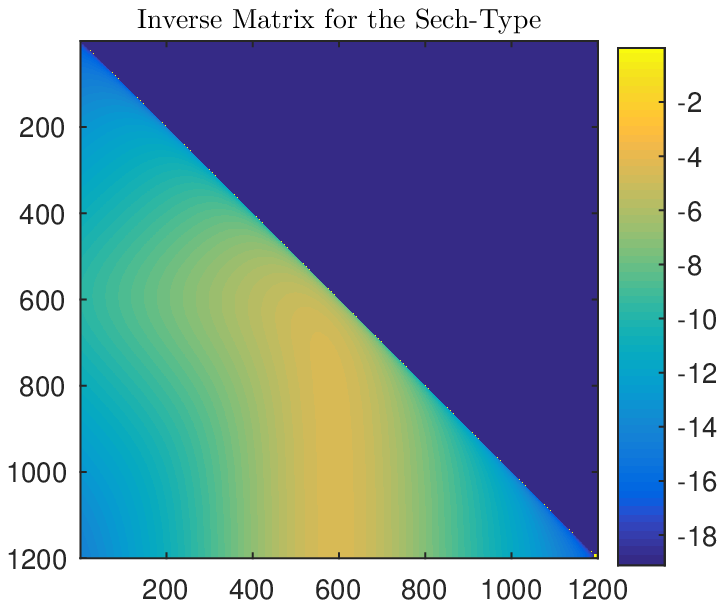}
\caption{(Color online) 2--D plot of $\mathrm{log}|1/(1-GVGV^{\ast})|$ for the initial wave packet in Eq.~(\ref{eq:sech}) for $\xi=0.0$.}
\label{F4}
\end{figure}

\begin{figure}
\includegraphics{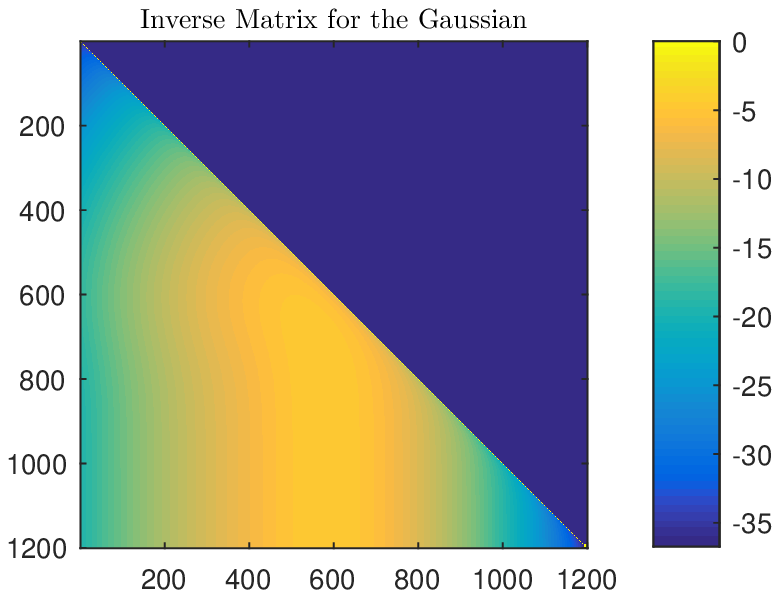}
\caption{(Color online) 2--D plot of $\mathrm{log}|1/(1-GVGV^{\ast})|$ for the initial wave packet given in Eq.~(\ref{eq:Gauss}) for $\xi=0.0$.}
\label{F5}
\end{figure}

\begin{figure}
\includegraphics{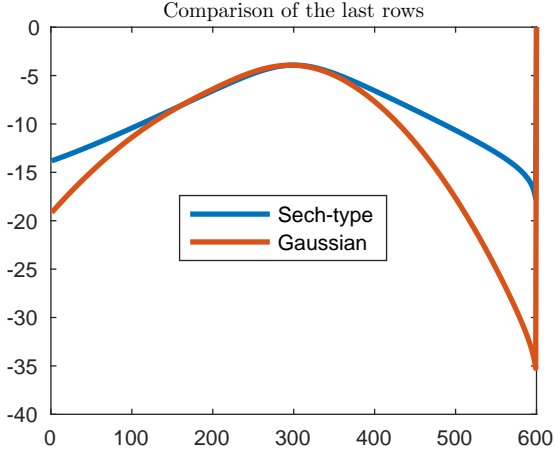}
\caption{(Color online) Simultaneous plot of the last rows of $\mathrm{log}|1/(1-GVGV^{\ast})|$ for the initial packets in Eq.~(\ref{eq:initWFs}) for $\xi=0.0$.}
\label{F6}
\end{figure}

The values of $|a(\xi)|$ for real $\xi$ under the initial conditions in Eqs.~(\ref{eq:sech}) and (\ref{eq:Gauss}) are shown in Fig.~\ref{F2}. As expected from the pure spliton condition, the transmission coefficient $a(\xi)$ for Eq.~(\ref{eq:sech}) is approximately unity and flat over a wide range of the spectral parameter, $0.0\leq\xi\leq 5.0$. By contrast, the plot of the transmission coefficient $a(\xi)$ for Eq.~(\ref{eq:Gauss}) is not a constant. This indicates that the Gauss--type wave packet can never become a pure soliton. In this simple example, our method successfully shows the well--known result, which has been given by analytic proof. In Fig.~\ref{F2}, $|a(\xi)|$ is smaller than unity. This is partially because of the truncating operation that we applied. What matters more is the insufficient value of $N$. We next confirmed the approach of $|a(\xi)|$ to unity when $N$ was increased. In Fig.~\ref{F3}, we show plots of $|a(\xi)|$ for various $N$, where all the plots were calculated for the same initial wave packet Eq.~(\ref{eq:sech}). 
In Figs.~\ref{F4} and \ref{F5}, visualized two--dimensional (2--D) plots of the matrix function $\mathrm{log}|1/(1-GVGV^{\ast})|$ for Eqs.~(\ref{eq:sech}) and (\ref{eq:Gauss}) under $\xi=0.0$ are respectively shown. These matrices are both lower--triangular. The last rows of them, which are used for the summing up, are simultaneously shown in Fig.~\ref{F6}. The sum of the last $N$--th row for Eq.~(\ref{eq:sech}) appears to take larger values around a narrow neighborhood of the origin $x=0.0$. For Eq.~(\ref{eq:Gauss}), the distribution is more widely spread. This reflects the feature of the pure soliton, that is, certain information is maximally concentrated. This kind of numerical experiment brings some insight that enables us to characterize the geometrical profile of pulse like functions.
\subsection{Double--peaked pulses}
Next, we consider some examples of double--peaked initial wave packets. This type of initial condition is given by a superposition of two wave packets, each of which has  a single peak that is appropriately apart from the other. There can be relative phases $\Delta \phi$ between these constituent wave packets. We show three typical examples of simulations for different  values of $\Delta\phi$.
Throughout this subsection, we set $N=600$ and the increment of $\xi$ to be 0.01 along the real and imaginary axes. 
\subsubsection{Case I: $\Delta\phi=0$}
\begin{figure}
\includegraphics{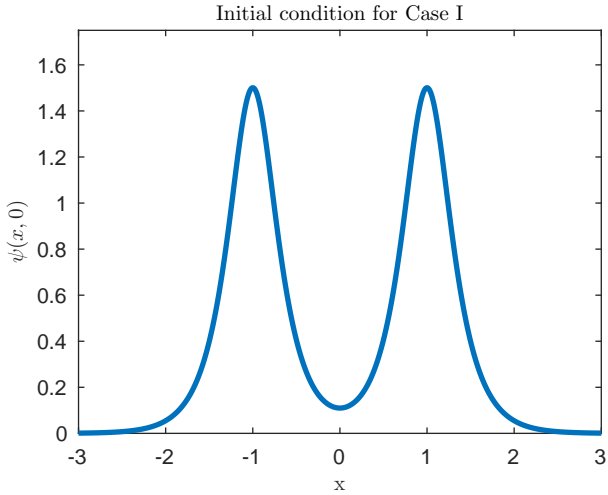}
\caption{(Color online) Initial wave packet for Case I. The profile of Eq.~(\ref{eq:32}) is depicted.}
\label{F7}
\end{figure}
\begin{figure}
\includegraphics{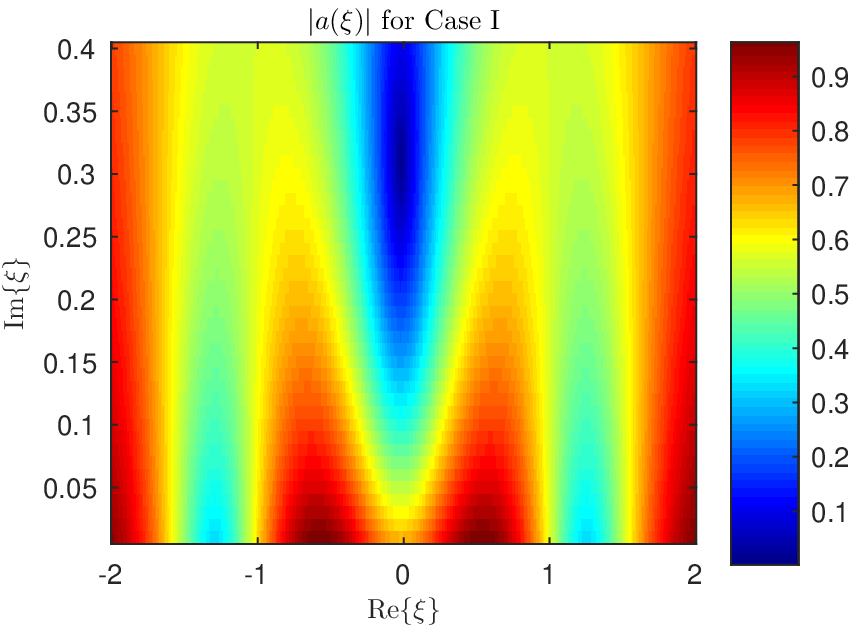}
\caption{(Color online) 2--D plot of $|a(\xi)|$ for Case I on the upper half plane of $\xi$.}
\label{F8}
\end{figure}
\begin{figure}
\includegraphics{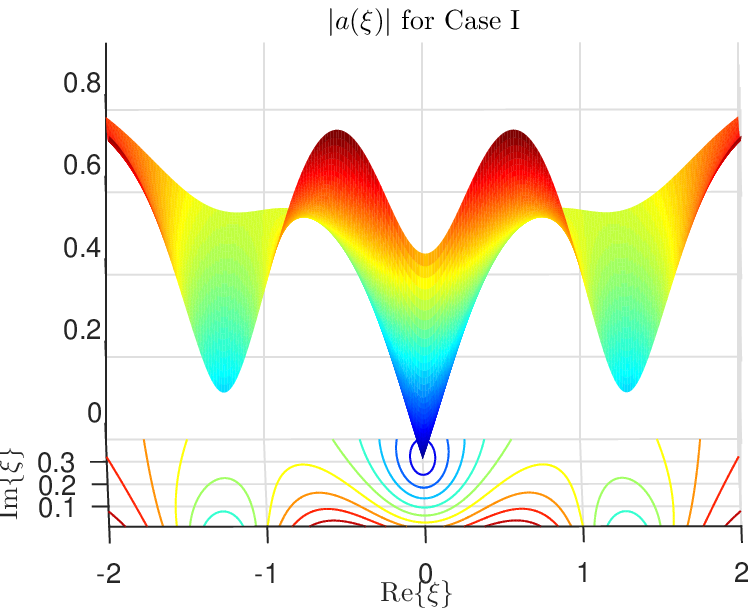}
\caption{(Color online) 3--D plot of $|a(\xi)|$ for Case I on the upper half plane of $\xi$.}
\label{F9}
\end{figure}
\begin{figure}
\includegraphics{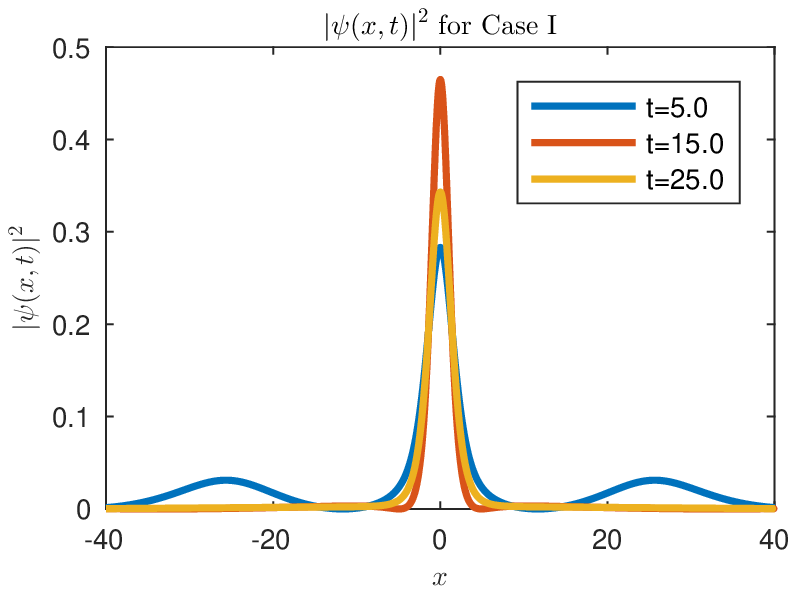}
\caption{(Color online) Absolute squares of the solution $\psi(x,t)$ for Case I at $t=5.0, 15.0$, and $25.0$.}
\label{F10}
\end{figure}
\begin{figure}
\includegraphics{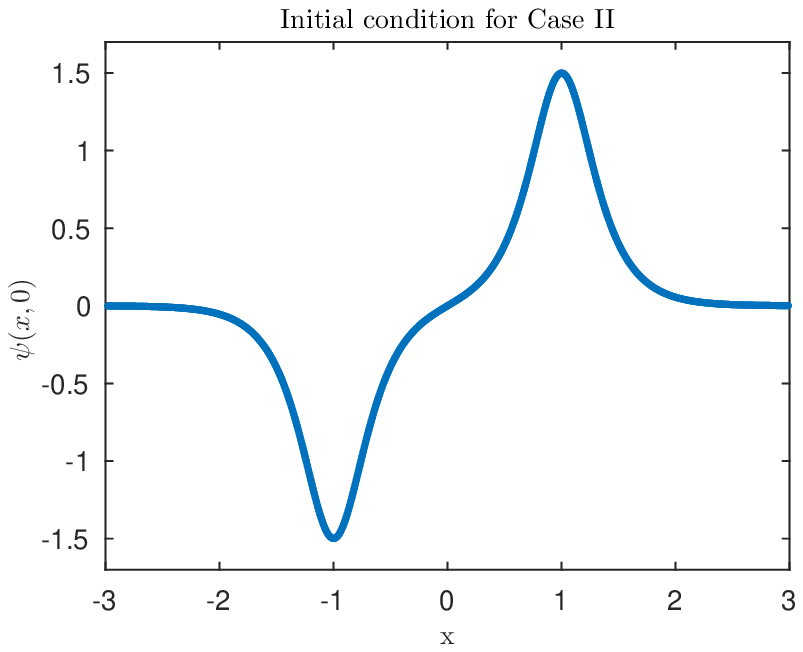}
\caption{(Color online) Initial wave packet for Case II. The profile of Eq.~(\ref{eq:33}) is depicted.}
\label{F11}
\end{figure}
\begin{figure}
\includegraphics{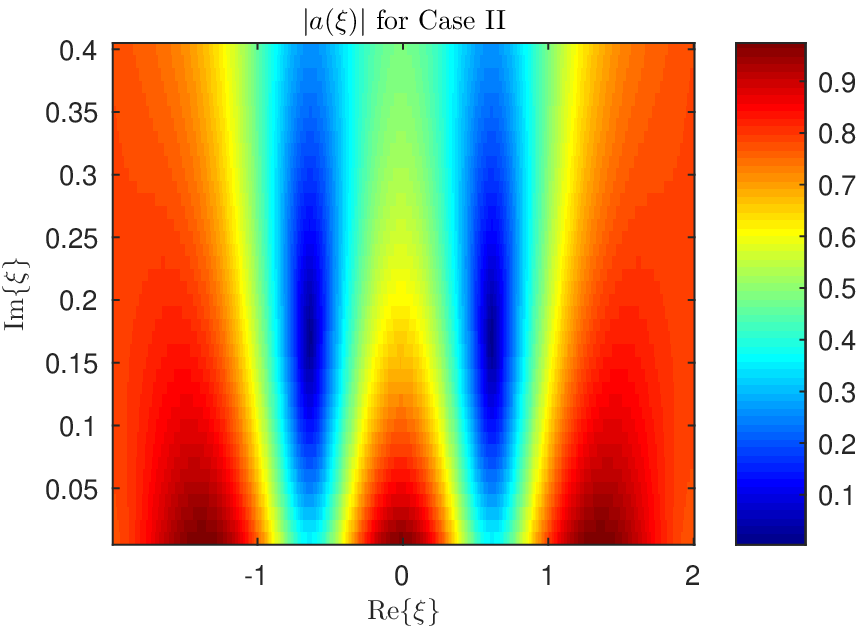}
\caption{(Color online) 2--D plot of $|a(\xi)|$ for Case II on the upper half plane of $\xi$.}
\label{F12}
\end{figure}
\begin{figure}
\includegraphics{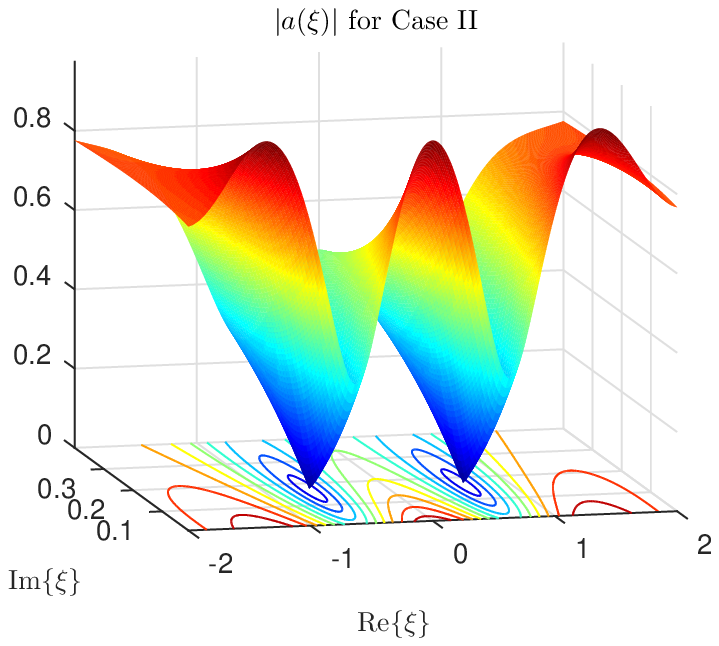}
\caption{(Color online) 3--D plot of $|a(\xi)|$ for Case II on the upper half plane of $\xi$.}
\label{F13}
\end{figure}
\begin{figure}
\includegraphics{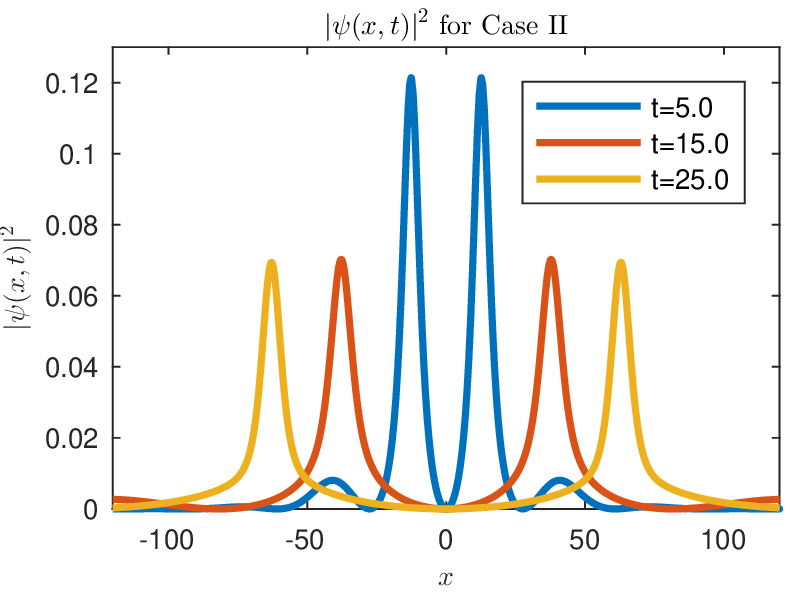}
\caption{(Color online) Absolute squares of the solution $\psi(x,t)$ for Case II at $t=5.0, 15.0$ and $25.0$.}
\label{F14}
\end{figure}
\begin{figure}
\includegraphics{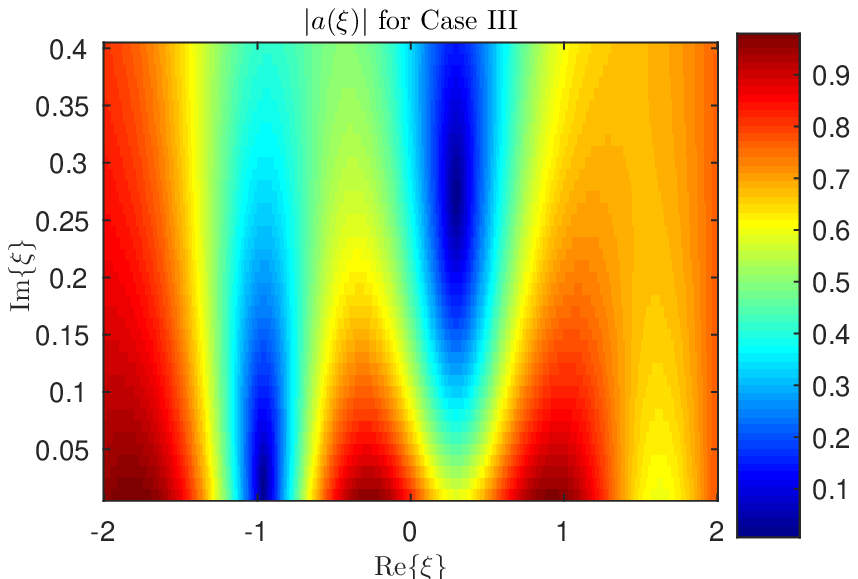}
\caption{(Color online) 2--D plot of $|a(\xi)|$ for Case III on the upper half plane of $\xi$.}
\label{F15}
\end{figure}
\begin{figure}
\includegraphics{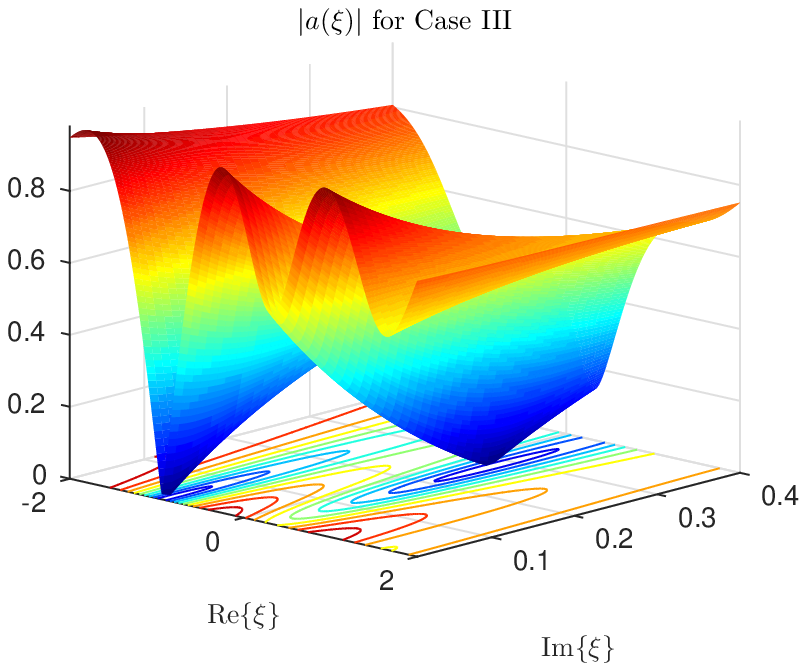}
\caption{(Color online) 3--D plot of $|a(\xi)|$ for Case III on the upper half plane of $\xi$.}
\label{F16}
\end{figure}
\begin{figure}
\includegraphics{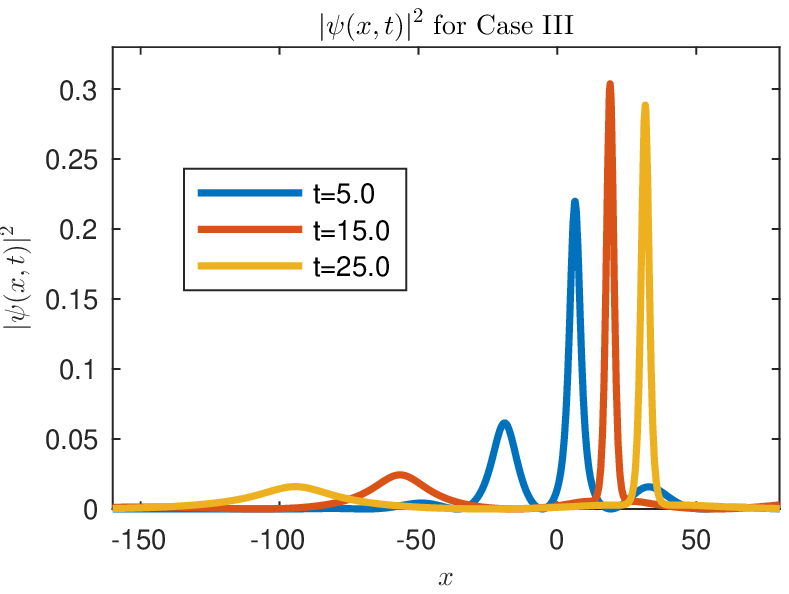}
\caption{(Color online) Absolute squares of the solution $\psi(x,t)$ for Case III at $t=5.0, 15.0$ and $25.0$.}
\label{F17}
\end{figure}
We consider the initial wave packet 
\begin{equation}
\psi(x,0)=1.5\left[\mathrm{sech}\{4(x-1.0)\}+\mathrm{sech}\{4(x+1.0)\}\right].
\label{eq:32}
\end{equation}
 No relative phase exists between the two constituent pulses in this case. The profile of this wave packet is shown in Fig.~\ref{F7}. We can see that it has two peaks at $x=\pm 1.0$. The absolute value $|a(\xi)|$ on the upper half plane of $\xi$ calculated by the method presented in the previous section is plotted in Fig.~\ref{F8}. The same quantity is also depicted three--dimensionally (3--D) in Fig.~\ref{F9}. The distribution is symmetric about the imaginary axis as expected, and a zero exists around $\xi=0.31i$. This means that only one soliton remains asymptotically. To verify this prediction, we numerically integrated the original NLSE. The absolute squares of the solution $\psi(x,t)$ at $t=5.0$, $15.0$, and $25.0$ are shown in Fig.~\ref{F10}, and we confirmed that only one pulse was generated around the origin.
\subsubsection{Case II: $\Delta\phi=\pi$}
We set the initial wave packet to be
\begin{equation}
\psi(x,0)=1.5\left[\mathrm{sech}\{4(x-1.0)\}-\mathrm{sech}\{4(x+1.0)\}\right].
\label{eq:33} 
\end{equation}
 This time, the relative phase $\Delta\phi$ is taken to be $\pi$. The profile of this wave packet is shown in Fig.~\ref{F11}. We can see that it has two peaks with opposite convexity at $x=\pm1.0$.
The absolute value of $a(\xi)$ for this case is plotted two--dimensionally in Fig.~\ref{F12} and three--dimensionally in Fig.~\ref{F13}. On the upper half plane of $\xi$, these plots are again symmetric about the imaginary axis, and two zeros appear around $\xi=\pm 0.62+0.17i$. This means that two solitons will remain in the final state. We confirmed this prediction by direct numerical integration. As shown in Fig.~\ref{F14}, we can observe two pulses symmetrically located around the origin as expected. The magnitudes of the two solitons are smaller than that of the single soliton generated in Case I. This is because the two pulses in Fig.~\ref{F14} are wider than the single pulse shown in Fig.~\ref{F10}. Actually, the heights of these two pulses remained at the level shown in Fig.~\ref{F14} even if we extended the termination time of the numerical integration. 
\subsubsection{Case III: $\Delta\phi=\frac{\pi}{2}$}
In this case, we take initial wave packet as 
\begin{equation}
\psi(x,0)=1.5\left[\mathrm{sech}\{4(x-1.0)\}+i\mathrm{sech}\{4(x+1.0)\}\right].
\end{equation}
The relative phase $\Delta\phi$ is set to $\pi/2$, which corresponds to the situation where one pulse approaches the other, then the two pulses collide. 
A 2--D plot of $|a(\xi)|$ is shown in Fig.~\ref{F15} and a 3--D plot of $|a(\xi)|$ is shown in Fig.~\ref{F16}. For this case, both plots are no longer symmetric. We can find two zeros of $a(\xi)$, one zero around $\xi=0.33+0.27i$ and the other around $\xi=0.93+0.01i$. For this initial condition, we also executed a numerical integration and observed two major pulses, a larger one and a smaller one, as shown in Fig.~\ref{F17}. Their centers are located asymmetrically about the origin.

In the above examples, the number of finally remaining solitons is at most two, but three or more solitons are generated if we choose suitable initial conditions. In fact, we showed that two box--type pulses with suitable height and separation develop to generate three solitons\cite{ref:Fujishima}. Concerning the number of solitons in the final state, the presented method has no limitations in principle, as long as $N$ is taken to be sufficiently large. We have confirmed that our previous result is also reproduced by the method presented in this paper.

An actual experiment on soliton collision has been performed, where attractively interacting BECs were exploited\cite{ref:Nguyen}. In the experiment, a confinement trap in the {\it x}--direction was applied and was not completely removed during the collision process. Owing to the existence of this trap, the geometry was not completely flat. The trapping potential is usually well--approximated by a quadratic function. It is known that the AKNS--ZS formulation for the NLSE can be extended to a situation where such a potential term exists\cite{ref:Balakrishnan}.

As long as we choose suitable initial conditions, by transforming a spatial variable {\it x} to a new one and reconstructing a new scattering potential from a given initial wave packet, we can use the same procedure as that demonstrated in previous sections. Nevertheless, it is unclear whether all initial conditions with a rapidly decreasing property work well in this extended scheme. There is a possibility that this method imposes additional restrictions on the class of initial conditions, which leads to a loss of generality. To avoid this complication, we neglected the effect of the trapping potentials in our illustration.

In the BEC experiment referred to above\cite{ref:Nguyen}, the effects of the relative phase on the collision process were examined and it was reported that the behaviors of the two condensate pulses were different in the in--phase and out--of--phase cases. To extract and discuss the role of the relative phase between condensates, our simplified models seem to be suitable and sufficient. In this sense, it can be reasonabley stated that these examples correspond to real physics and have physical interest.

\section{Summary and Discussion}
In this paper, we have presented a novel approximation method for obtaining the scattering data for the ZS problem reduced from the one--dimensional NLSE under general rapidly decreasing initial conditions. Using the obtained scattering data, the numbers of asymptotically generated solitons have been estimated for initial conditions whose profiles are not limited to pure solitons. The derived results are consistent with those of direct numerical integration. These results include brief models for a collision experiment on BEC solitons. A similar experiment on soliton collision in a homogeneous space is expected to be more easily conducted in the field of nonlinear optics. 
Because of its ubiquitous appearance and importance in physics, we have limited our illustration to the NLSE. The method developed in this paper is essentially applicable to a wide class of soliton equations that can be formulated as the AKNS, Kaup--Newell,\cite{ref:KN} and Wadati--Konno--Ichikawa\cite{ref:WKI} systems. Other soliton equations of importance that can be fit into this format include the Korteweg--de Vries (KdV), modified KdV, sine--Gordon, and derivative NLS equations. 

Since the NLSE is a Hamilton system, an action variable $J(\xi)$ and an angular variable $\theta(\xi)$ exist, with which the Hamiltonian can be written. Excluding the zeros, they are expressed by the scattering data as:
\begin{equation}
J(\xi)=-\frac{1}{\pi}\mathrm{log}a(\xi)a^{\ast}(\xi),
\end{equation}
\begin{equation}
\theta(\xi)=-\frac{1}{2i}\mathrm{log}\frac{b(\xi)}{b^{\ast}(\xi)}.
\end{equation}  
Using the technique developed in this paper, we have obtained a new numerial method of investigating the geometric surfaces defined by these action--angle variables. Although we have shown only the absolute values $|a(\xi)|$ for visualization, the original complex geometry of the scattering data in the upper half plane of $\xi$ is expected to include richer information. 
Our ultimate goal is to find a way to extract physical information through the investigation of these geometrical entities. It is desirable that we can define some characteristic quantities concerning the geometry so that they correspond to actual physical information. This task is left as future work.

\begin{acknowledgment}
The authors express their sincere gratitude to Professor Ralph Willox of the University of Tokyo and Professor Ken--ichi Maruno of Waseda University for their interest in this work and stimulating discussions. One of the authors (H.~F.) thanks Utsunomiya University for offering opportunities for fruitful discussions.
\end{acknowledgment}

\end{document}